\documentclass[openacc]{rstransa}

\providecommand{\f}[2]{\frac{{#1}}{{#2}}}
\newcommand{\ee}[1]{\begin{equation}#1\end{equation}}
\newcommand{\ea}[1]{\begin{align}#1\end{align}}

\begin{document}

\title{Vacuum Stability in the Early Universe and the Backreaction of Classical Gravity}

\author{
Tommi Markkanen}

\address{Department of Physics, King's College London, Strand, London WC2R 2LS, United Kingdom\\
}

\subject{Particle cosmology}

\keywords{Particle-theory and field-theory models of the early Universe, Quantum fields in curved spacetime}

\corres{Tommi Markkanen\\
\email{tommi.markkanen@kcl.ac.uk}}

\begin{abstract}
In the case of a metastable electroweak vacuum the quantum corrected effective potential plays a crucial role in the potential instability of the Standard Model. In the Early Universe, in particular during inflation and reheating, this instability can be triggered leading to catastrophic vacuum decay. We discuss how the large spacetime curvature of the Early Universe can be incorporated in the calculation and in many cases significantly modify the flat space prediction. The two key new elements are the unavoidable generation of the non-minimal coupling between the Higgs field and the scalar curvature of gravity and a curvature induced contribution to the running of the constants. For the minimal set up of the Standard Model and a decoupled inflation sector we show how a metastable vacuum can lead to very tight bounds for the non-minimal coupling. We also discuss a novel and very much related dark matter generation mechanism.
\end{abstract}


\begin{fmtext}
\section{Introduction}
The discovery of the Higgs boson has been the most significant discovery of the LHC and marks the verification of the last constituent of the Standard Model of particle physics.  However, the Standard Model does not seem to provide explanations for some of the deep mysteries in physics such as baryon asymmetry, neutrino masses or dark matter. Unfortunately, there has been little experimental results from particle accelerators to guide theoretical work.

One of the most surprising realizations of recent years has been the potential instability of the electroweak vacuum \cite{Buttazzo:2013uya,Degrassi:2012ry}: the current best fit observations imply the striking feature that the Standard Model effective potential of the Higgs field will at large scales generate a second minimum, sometimes referred as the true \end{fmtext} \maketitle \noindent vacuum, with substantially negative energy-density. The generation of a second minimum is ultimately a quantum effect resulting from the energy scale dependence, or running, of the parameters. The most accurate calculations \cite{Buttazzo:2013uya,Bednyakov:2015sca,Degrassi:2012ry} are converging towards the result that the energy scale $\Lambda_I$ at which the effective, or quantum corrected, potential turns over to negative values lies between $10^{10}$ -- $10^{12}$GeV\,, with an absolutely stable potential located within the 3$\sigma$ bound around the central values. The emergence of a second minimum unavoidably implies that the current electroweak vacuum has a finite lifetime and thus will eventually decay. If the lifetime of the current vacuum is sufficiently long, this not a contradiction with current observations, however crucially, during and epoch of cosmological inflation the situation can change: fluctuations of the Higgs field are amplified by the expansion of space during inflation, in many ways analogously to a thermal bath, and for large enough fluctuations there in fact exists a dangerously large probability that the electroweak vacuum decays during inflation \cite{Hook:2014uia,Fairbairn:2014zia}\footnote{See \cite{Markkanen:2017dlc} for a more complete list of related work}. 

From the non-observation of primordial tensor modes the combined BICEP2/Keck and Planck data points the bound of the scale of inflation to be $H\sim10^{14}$GeV or lower \cite{Array:2015xqh}. As an order of magnitude leading approximation, the probability density of vacuum decay during inflation scales as 
\begin{equation}
P\sim \exp\left\{-\f{8\pi^2V_{\rm max}}{3H^4}\right \}\, ,\label{eq:p}
\end{equation}
where $V_{\rm max}$ refers to the maximum of the effective potential. This then leads to the approximate criterion
\ee{V_{\rm max}\gtrsim H\,,\label{eq:p0} }
for the survival for the electroweak vacuum during inflation assuming a metastable Standard Model.
 Any theory with a decay to the true vacuum would likely not give rise to the Universe we observe and as such the requirement of inflationary vacuum stability results in a consistency constraint linking cosmological history to particle physics: the prediction of vacuum decay in the Early Universe would be a cosmological verification for the need of beyond the Standard Model physics. From (\ref{eq:p}) one may also see that any mechanism increasing the barrier between the electroweak and the true vacuum exponentially decreases the vacuum decay probability. 

In \cite{Herranen:2014cua,Herranen:2015ima} it was realized that gravitys presence in the quantum dynamics of the Early Universe cannot be neglected: when consistently including the backreaction of gravity, a metastable electroweak vacuum can be compatible with large scale inflation, as long as the non-minimal coupling of the Higgs boson to gravity is of a sufficient size. In this work we review the details of the derivation in \cite{Herranen:2014cua,Herranen:2015ima} and furthermore discuss how a related mechanism can lead to an efficient production of dark matter \cite{Markkanen:2015xuw}. 

\section{Random walk during inflation}

If inflation is given by some yet unknown physics and not the Higgs field itself as in \cite{Bezrukov:2007ep}, on an exponentially expanding background the Higgs field $\phi$ behaves as a stochastic spectator field, whose probability distribution $ {P}(t,\phi)$ may be calculated from the Fokker-Planck equation \cite{Starobinsky:1994bd}
\begin{equation}
\dot{P}(t,\phi)=\frac{1}{3H}\frac{\partial}{\partial \phi}\big[ {P}(t,\phi){V}'(\phi)
\big]+\f{H^3}{8\pi^2}\frac{\partial^2}{\partial \phi^2 }{P}(t,\phi)\,,\label{eq:FP}
\end{equation}
where $V(\phi)$ is the classical potential. The essential assumption that leads to the stochastic description is treating the ultraviolet physics as a white noise contribution which induces "random walk" in the long wave length modes. 

By only invoking the stochastic approach for the vacuum stability/instability analysis we are neglecting the bubble nucleation transition "through the barrier" via the Coleman--de Luccia instanton. However, vacuum decay via bubble nucleation is subdominant with respect to the stochastic fluctuations and because of this to a leading approximation can be neglected \cite{Hook:2014uia}. The approximation in (\ref{eq:p}) follows from (\ref{eq:FP}) by solving for the stationary case $\dot{P}(t,\phi)=0$ and determining the probability of a "jump" high enough to reach the top of the potential.

If the scale of inflation $H$ is large one may solve the variance of a light self-interacting scalar field with $V(\phi)\sim \lambda \phi^4$, such as the Higgs to acquire substantial long wave length fluctuations as
\ee{\langle \phi^2 \rangle\sim \f{H^2}{\sqrt{\lambda}}\,,}\
indicating that a dangerous fluctuation may be generated for a sufficiently large $H$. But however convenient the stochastic approach may be, when studying the cosmological implications from the electroweak vacuum instability we must introduce an important modification to the above prescription: since the reason why the Standard Model vacuum is unstable is intimately linked to renormalization and hence ultraviolet physics, it is not visible when using the classical potential $V(\phi)$ in equation (\ref{eq:FP}). Hence, we need to replace $V(\phi)$ with the effective potential that is sensitive to the quantum effects, which we denote with $V_{\rm eff}(\phi)$, in (\ref{eq:FP}). The important point here is that the infra-red modes are correctly taken into account by the Fokker-Planck equation (\ref{eq:FP}) so $V_{\rm eff}(\phi)$ should include only the ultraviolet physics. Furthermore, since we are interested in physics during inflation with a large scale $H\gg \Lambda_I$, the curvature of the background must be included in the computation of $V_{\rm eff}(\phi)$. 

\section{Effective potential in curved space}\label{subsec:prod}
\begin{table}
\caption{\label{tab:contributions}The effective potential (\ref{potential}) with $W^{\pm}$, $Z^0$, top quark t, Higgs $\phi$ and the Goldstone bosons $\chi_{1,2,3}$. The non-trivial nature of the curvature corrections is apparent in the coefficients $\theta_i$.}
\vspace{2mm}
\begin{center}
 \begin{tabular}{|c||cccccc|}
 \hline
   $\Phi$ & $~~i$ & $~~n_i$  &$~~\kappa_i$ & $\kappa'_i$          & $~~\theta_i$   & $\quad c_i~~$ \\[1mm]\hline
   $~$ & $~~1$  & $~~2$       & $~~ g^2/4$        & $0$        & $~~{1}/{12}$     & $\quad{3}/{2}~~$ \\[1mm]
   $~W^\pm$ & $~~2$  & $~~6$       & $~~ g^2/4$        & $0$        & $~~{1}/{12}$      & $\quad{5}/{6}~~$ \\[1mm]
   $~$ & $~~3$  & $-2$      & $~~g^2/4$        & $0$        & $-{1}/{6}$      & $\quad{3}/{2}~~$ \\[1mm]\hline
   $~$ & $~~4$  & $~~1$       & $~~(g^2+g'^2)/4$ & $0$        & $~~{1}/{12}$     & $\quad{3}/{2}~~$ \\[1mm]
   $Z^0$ & $~~5$  & $~~3$       & $~~(g^2+g'^2)/4$ & $0$        & $~~{1}/{12}$      & $\quad{5}/{6}~~$ \\[1mm]
   $~$ & $~~6$  & $-1$      & $~~(g^2+g'^2)/4$ & $0$        & $-{1}/{6}$     & $\quad{3}/{2}~~$ \\[1mm]\hline
   t & $~~7$  & $-12$     & $~~ y_{\rm t}^2/2$      & $0$        & $~~{1}/{12}$     & $\quad{3}/{2}~~$ \\[1mm]\hline
   $\phi$ & $~~8$  & $~~1$       & $~~3\lambda$           & $~m^2$      & $~~\xi-{1}/{6}$  & $\quad{3}/{2}~~$
\\[1mm]\hline
   $\chi_i$ & $~~9$  & $~~3$       & $~~\lambda$            & $~m^2$      & $~~\xi-{1}/{6}$   & $\quad{3}/{2}~~$
\\[.5mm]\hline
  \end{tabular}
  \end{center}
  \end{table}
The calculation for the quantum corrected effective potential at the ultraviolet limit is surprisingly simple when compared to what one usually encounters when deriving an effective potential on a curved background. The reason for this is that at the ultraviolet limit all well-behaved field theories must have universal behaviour, which translates as the fact that the divergences that are generated are independent of the state in which they are calculated. Hence, at very high momenta even with a background with large curvature the effective potential resembles very much the well-known flat space results. Another way of understanding the simplification arising at short wave-length modes is realizing that any smooth surface when magnified enough will approach flat space. 

The effective potential in the high ultraviolet on an arbitrary curved background we can very conveniently derive via the resummed Heat Kernel method \cite{Jack:1985mw}. This includes the essential curvature terms. We point out that this derivation is completely semi-classical or that gravity is not quantized. Since $\Lambda_I/M_{\rm pl}\ll1$ this is an accurate approximation when the Higgs is not the dominant energy component, as was recently rigorously shown to be true in \cite{Markkanen:2017dlc}.

In what follows we have chosen a minimal set-up where there are no direct couplings between the inflaton sector and the Standard Model. In terms of radiative stability this is consistent: no couplings will be generated by renormalization group running if the tree-level values vanish. Having a direct coupling to the inflaton sector as many reheating models require would have an impact on the vacuum stability. In such cases the results become very much model dependent. However, our main arguments related to the necessary generation of a non-minimal coupling to gravity are independent of possible couplings to the inflaton sector.

In strict de Sitter space to 1-loop order the result in the 't Hooft-Landau gauge including only the relevant degrees of freedom for the quantum corrected, or effective, potential was first derived in \cite{Herranen:2014cua}
\begin{align}
V_{\rm eff}(\phi)& = -\f{1}{2}m^2\phi^2 + \f{1}{2}\xi R\phi^2 + \f{1}{4}\lambda\phi^4
+ \sum\limits_{i=1}^9 \f{n_i}{64\pi^2}M_i^4(\phi)\left[\log\f{\big|M_i^2(\phi)\big|}{\mu^2} - c_i \right]\,;\label{potential}\\M^2_i(\phi)&=\kappa_i\phi^2-\kappa'_i +\theta_i R\, ,\label{eq:effm}
\end{align}
with $R=12H^2$ and where the $n_i$ represent the various degrees of freedom described in table \ref{tab:contributions} with  $M_i(\phi)$ as their effective masses. To be specific, the field $\phi$ is precisely the scalar degree of freedom of the Higgs doublet which by developing an expectation value at low energies gives rise to the masses for the Standard Model constituents.

When deriving (\ref{potential}) we have not included all contributions from curvature: generically at the level of the action divergences proportional to $R_{\mu\nu}R^{\mu\nu}$ and $R_{\mu\nu\eta\rho}R^{\mu\nu\eta\rho}$ are also generated  \cite{Birrell:1982ix}. This however does not result into any new $\phi$-dependence at tree-level and only a mild $\phi$-dependence will be generated through $\phi$-dependent logarithms, which here we neglect.

The potential in  (\ref{potential}) can be seen to resemble very much the 1-loop flat space result \cite{Casas:1994qy} for the reasons given at the beginning of this section. Precisely as in the flat space derivation, (\ref{potential}) is not applicable for the scales that are relevant for the vacuum instability $\phi \geq\Lambda_I$. The reason for this lies in $\mu$, the renormalization scale: the parameters are matched to observables at scales several oders of magnitude smaller than $\Lambda_I$, which at high energies leads to large logrithms and indicates a breakdown of the perturbative expansion. The well-known technique of renormalization group improvement may then be used to rectify the situation, which gives a potential where coupling constants run according to the energy scale one is probing. At its core, renormalization group improvement is a statement of the independence of physics to the renormalization scale, so the improved result must satisfy ${d}V_{\rm eff}/{d\mu}=0$. For a perturbative result however, the truncated higher order contributions always contain some left-over $\mu$-dependence, and one should make a choice for $\mu$ such that the convergence of the improved effective potential is optimized \cite{Ford:1992mv}, essentially this comes about by making sure that the logarithms remain small. In flat space for $\phi\gg m$ a good and frequently used choice is $\mu=\phi$, which as its leading approximation gives the frequently used result \ee{V_{\rm eff}(\phi)\approx\f{\lambda(\phi)}{4}\phi^4\,;\qquad H=0\,.\label{flat}}

When the background is strongly curved the same prescription of keeping the logarithms under control forces one to make a distinctively different choice. As one may see from the effective potential (\ref{potential}), in de Sitter space all effective masses contain curvature contributions, remembering that the scalar curvature in de Sitter space is $R=12H^2$. 
A well-behaved choice in curved space can then be obtained from
\begin{equation}
\mu^2=\phi^2+R\,.\label{eq:mu}
\end{equation} The fact that the renormalization group running scale $\mu$ gets a contribution from $R$  leads to an important result: curvature can influence the running of couplings. 

A related and equally important curved space effect to curvature induced running is the generation of the non-minimal term which couples the scalar curvature to the Higgs field $\sim \xi R \phi^2$. The expression for the effective potential (\ref{potential}) shows that we chose a non-minimal term to be present already at tree-level, but from the quantum correction one may see that this is in fact required by renormalization group running: the $\beta$-function for $\xi$ can be determined from the coefficients of the logarithms to be \ee{
16\pi^2 \beta_{\xi} = \Big(\xi - \f{1}{6}\Big)\left(12\lambda + 6y_{\rm t}^2 - \f{3}{2}g'^2 - \f{9}{2}g^2\right)\,,
\label{betaxi}
}  from which it is apparent that $\xi=0$ is not an fixed point of the renormalization group flow. From this we can conclude that a non-zero $\xi$ will always be generated by a change in the energy scale, if from nothing else than from the inevitable change of the Hubble rate $H$ during the evolution of the Universe. 

The two modifications introduced by background curvature are most clearly visible when using the approximation to the 1-loop renormalization group improved effective potential at a large scale ($\phi\gg m$) with a tree-level potential but with running couplings
\begin{equation}
V_{\rm eff}=\f{\xi(\mu)}{2}R\phi^2+\f{\lambda(\mu)}{4}\phi^4\, ,\label{eq:lo}
\end{equation}
where we have chosen $\mu$ as in (\ref{eq:mu}). The effects can be easily understood by comparing the flat space result (\ref{flat}) to (\ref{eq:lo}) and we list them here once more for clarity:

\begin{itemize}
\item[(1)] \textit{Spacetime curvature induces running of couplings.\\} 

\item[(2)]\textit{A non-zero $\xi$ is unavoidable.}
 
\end{itemize}
\begin{figure}
\begin{center}
\hspace{-0.5cm}\includegraphics[width=0.7\textwidth]{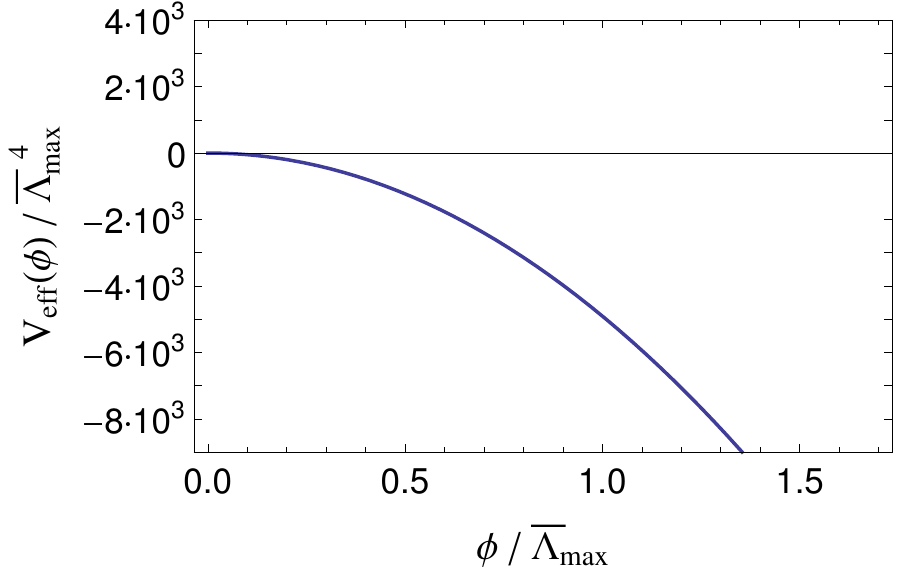}
\caption{\label{ff}The behaviour of the renormalization group improved 1-loop effective potential, with the Hubble rate $H=10^{10}$GeV and the choice $\xi=0$ at the electroweak scale. The normalization scale $\overline{\Lambda}_{\rm max}$ refers to the scale at which the flat space maximum is reached. Due to the generation of a negative curvature mass term the potential is monotonically decreasing.}
\end{center}
\end{figure}
\section{Stability during inflation}
We now have everything we need for studying the stability of the electroweak vacuum during inflation. The main features may be understood from the tree-level improved result (\ref{eq:lo}). 

\begin{figure}
\begin{center}
\hspace{-0.5cm}\includegraphics[width=0.7\textwidth]{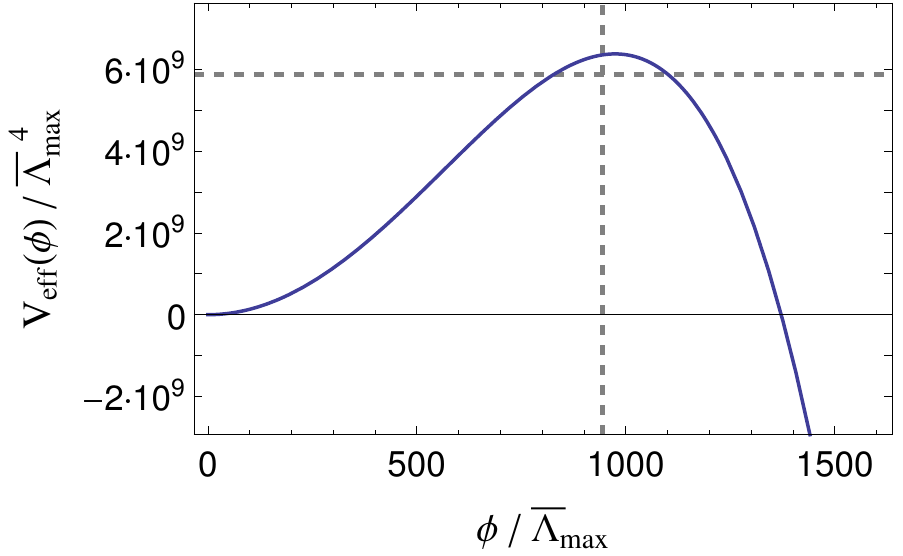}

\caption{\label{f}The behaviour of the renormalization group improved 1-loop effective potential, with the Hubble rate $H=10^{10}$GeV and the choice $\xi=0.1$ at the electroweak scale. The normalization scale $\overline{\Lambda}_{\rm max}$ refers to the scale at which the flat space maximum is reached, indicating that here the peak occurs at a significantly higher scale.}
\end{center}
\end{figure}

Due to the curvature induced running we discussed in the previous section, the four-point coupling for the Higgs field $\lambda(\mu)$ is negative from the very onset of inflation, even if $\phi=0$, if the scale of inflation $H$ is larger than the instability scale $\Lambda_I$. From this we can deduce that for large inflationary scales the only means of having a positive potential is a large non-minimal term $(1/2)\xi R \phi^2$ that counteracts the negative contribution from the quartic term in (\ref{eq:lo}). If $\xi$ is renormalized to have a vanishing value at the electroweak scale it generates a negative value when evaluated at a high scale \cite{Herranen:2014cua}, which is shown in figure \ref{ff}. However, as we explained in the previous section, since $\xi$ is not a fixed point in the renormalization group running $\xi=0$ is not radiatively stable. Furthermore the current observable bound for $\xi$ is for all practical purposes non-existent \cite{Atkins:2012yn} so most non-zero choices for $\xi$ are equally motivated physically. Already when making relatively modest choices for $\xi$, for example $\sim 0.1$ at electroweak scales, the non-minimal term runs to give a large positive contribution at high energies as visible in figure \ref{f}. 

As a first approximation since the couplings run quite weakly at large scales, we can roughly find the maximum of the potential when $\lambda$ is negative but $\xi$ positive by using (\ref{eq:lo}):
\begin{equation}
\Lambda^2_{\rm max}\approx-\frac{\xi(\mu) }{\lambda(\mu)}R\,,\qquad V_{\rm max}\approx-\frac{\left[\xi(\mu)R\right]^2}{4\lambda(\mu)}\,;\qquad \mu^2\approx R\, ,\label{eq:anapp}
\end{equation}
which are represented by the dashed lines in figure \ref{f}. This verifies the statement that for $\xi$ that is $\mathcal{O}(1)$ or less at electroweak scales we already have $V_{\rm max}\geq H^4$ and the scaling in (\ref{eq:p}) shows that the probability of vacuum decay is diminished significantly. This can also be seen in figure \ref{f}, where we plot the full RG improved potential in curved space to 1-loop accuracy. In figure \ref{f} we have also normalized the $x$-axis with respect to the field value at which the flat space potential reaches its maximum, $\overline{\Lambda}_{\rm max}$. As the representative scale for inflation we have chosen $H\sim10^{10}$GeV, since in the 1-loop approximation $\lambda$ runs to negative values at much smaller scales than in a the state-of-the-art results \cite{Buttazzo:2013uya}, roughly at $\Lambda_I\sim10^8$GeV. Hence our choice corresponds to $H\gg\Lambda_I$ with respect to the 1-loop result. 

Clearly, from figure \ref{f} we see that the peak of the potential is reached at a scale that is $\sim 10^{3}$-times larger than $\overline{\Lambda}_{\rm max}$. Importantly, the maximum of the potential is correspondingly increased and from (\ref{eq:anapp}) we see it scaling roughly as $V^{1/4}_{\rm max}\sim 2H$. Formula (\ref{eq:p}) then gives the order of magnitude estimate for the transition probability to the unstable vacuum as $P<e^{-400}$, showing that for $\xi \gtrsim0.1$ fixed at the electroweak scale the potential is stable at high energies. 

All values for $\xi$ that are larger than some threshold will lead to a similar stabilizing result. A special point is reached at $\xi=1/6$ after which the Higgs starts behaving as a non-fluctuating massive field and the instability problem is completely removed. 
\begin{figure}
\begin{center}
\hspace{-0.5cm}\includegraphics[width=0.75\textwidth]{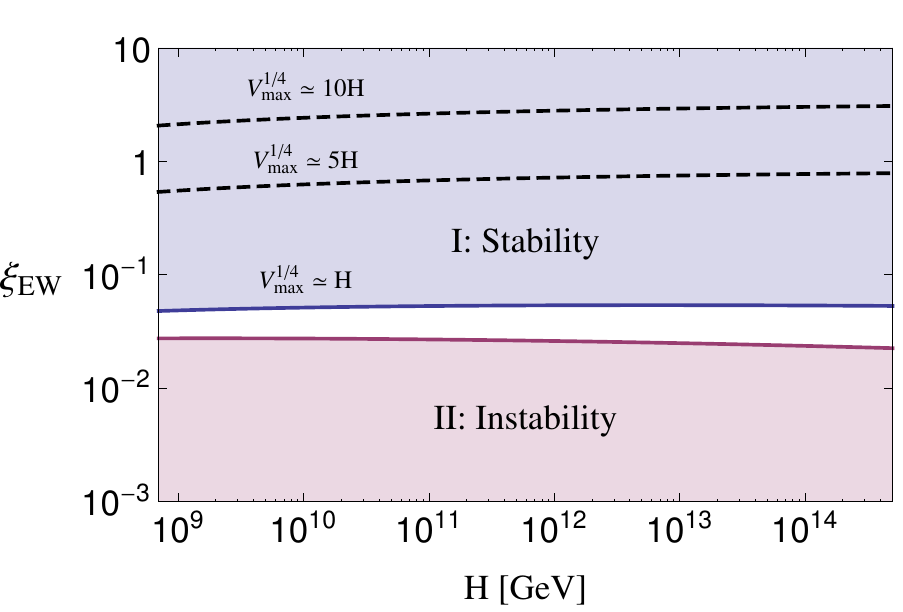}
\end{center}
\caption{\label{f2}The regions I (blue, top) for which $V_{\rm max}^{1/4} \gtrsim H$ leading to a suppressed transition probability
 to the true vacuum, and II (red, bottom), where the EW vacuum is unstable due to a monotonically decreasing potential from a negative curvature contribution.}
\end{figure}
As the order of magnitude criteria for stability we can use (\ref{eq:p0}) and for the instability the point when the potential becomes monotonically decreasing. The results for various initial values of $\xi$ at the electroweak scale may be found in figure \ref{f2}, which show that the Standard Model is compatible with large scale inflation as long as $\xi\gtrsim 10^{-2}$ at the electroweak scale.
\section{Stability after inflation}
The fact that the Standard Model may be stable during inflation even when $H$ is as large as allowed by the tensor bound does not mean that a vacuum decay is completely avoided in the Early Universe. This is due to the explosive and non-perturbative dynamics which often occur after inflation has ended and the thermal plasma of the hot Big Bang becomes the dominant energy component of the Universe. This is of course what happens during the reheating epoch.

After the end of inflation a generic feature of many inflationary models is the coherently oscillating inflaton field around the minimum of its potential. This can give rise to a very potent non-perturbative amplification of resonant quantum modes dubbed preheating \cite{Kofman:1997yn}. Right after inflation before any notable thermalization has occurred there is a significant possibility that a large enough fluctuation is generated by preheating to induce vacuum decay, which was first noticed in \cite{Herranen:2015ima}. The decay, quite surprisingly, does not require a direct coupling between the Higgs field and the inflaton but can result from the non-minimal term $\sim \xi R\phi^2$.

\begin{figure}
\begin{center}
\includegraphics[width=0.5\textwidth]{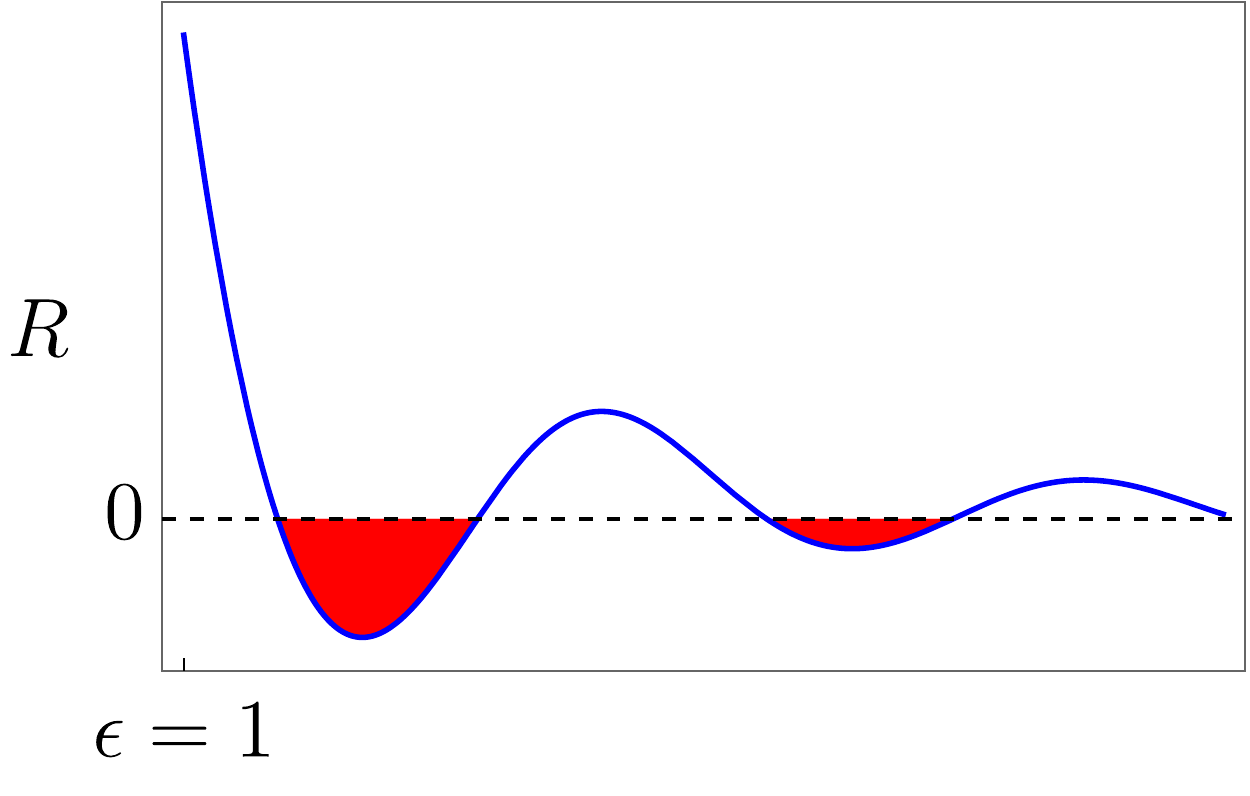}
\caption{\label{reh} The behaviour of the scalar curvature $R$ for $m_\Phi^2\Phi^2$-inflation. Due to the expansion of space the first oscillation gives rise to the largest tachyonic amplification. $\epsilon\equiv -\dot{H}/H^2=1$ signals the end of inflation}
\end{center}
\end{figure}

Let us consider a generic inflaton potential $V_{\rm inf}(\Phi)$ during the coherent oscillations of an inflaton field $\Phi$ around the origin of $V_{\rm inf}(\Phi)$. By our assumptions, $\Phi$ is the dominant energy component and we can solve the behaviour of $R$ by taking the trace of the Einstein equation
\ee{M_{\rm pl}^2G_{\mu\nu}=T_{\mu\nu}\qquad\Rightarrow\qquad R=\f{1}{M_{\rm pl}^{2}}\left[4 V_{\rm inf}(\Phi)-\dot{\Phi}^2\right]\,.\label{R}}
Assuming further that the minimum of $V_{\rm inf}(\Phi)$ is at $\Phi=0$, when a coherently oscillating $\Phi$ crosses the minimum of its potential from equation (\ref{R}) we see that $R<0$. This means that periodically the non-minimal term $\sim \xi R\phi^2$ gives rise to an imaginary mass term for the Higgs field. A negative mass squared will in turn give rise to extremely efficient particle creation \cite{Dufaux:2006ee} where the particle number increases much faster than from the resonant effects usually encountered in preheating. Because of this even during the first half of the first oscillation a significant fluctuation for the Higgs field can be generated. In figure \ref{reh} we show this effect for inflation with the potential $V_{\rm inf}(\Phi)\sim m_\Phi^2\Phi^2$.

Since the effective mass is periodically imaginary, the amplification from the $\xi R$-term may be viewed as tachyonic amplification, which was first discussed in  \cite{Bassett:1997az,Tsujikawa:1999jh}. By making the canonical reheating approximation where we assume the oscillations of the inflaton to be sinusoidal, the amplification of a quantum mode $f(t)$ of the Higgs field
\ee{\hat{\phi}=\int \f{d^3\mathbf{k}}{\sqrt{(2\pi)^3}}\Big[\hat{a}_\mathbf{k}f(t)\,e^{-i \mathbf{k}\cdot\mathbf{x}}+{\rm H.C}\Big]} 
for $m_\Phi^2\Phi^2$-inflation can be expressed as the Mathieu equation \cite{Bassett:1997az,Herranen:2015ima}
\ee{\f{d^2f(t)}{dz^2}+\bigg[A_k-2q\cos(2z)\bigg]f(t)=0,\qquad z=m_\Phi t\,,\label{eq:mathieu}}
\ee{\nonumber A_k= \f{\mathbf{k}^2}{a^2m_\Phi^2}+\xi \f{\Phi^2}{2 M_{\rm pl}^2},\qquad q=\f{3\Phi^2}{4 M_{\rm pl}^2}\bigg(\f{1}{4}-\xi\bigg)\, .}
With the help of the analysis of \cite{Dufaux:2006ee} we can derive a lower bound for the occupation number after the first oscillation
\ee{n^{1}_\mathbf{k}=e^{2X_\mathbf{k}}\,,\qquad X_\mathbf{k} \approx \sqrt{\xi}\f{\Phi}{M_{\rm pl}}\, \label{eq:occapp},}
which is directly related to the generation of a variance and hence a large fluctuation as
\ee{\langle\hat{\phi}^2\rangle\sim \int d^3\mathbf{k}|f(t)|^2\,n^{1}_\mathbf{k}\,,}
where to be conservative we have only included the superhorizon modes. Hence the amplification is exponential with the opposite behaviour to the inflationary case discussed in the previous section i.e. the larger the $\xi$ the stronger the effect leading to a larger vacuum decay probability for $\xi \gg 1$.

Before we can claim that the non-minimal coupling for the Higgs field can give rise to vacuum decay during reheating, we must analyse how the backreaction from the   created particles modifies the tachyonic resonance. Indeed, it is a generic feature of preheating that once the particle density resulting from the resonance becomes significant its backreaction will lead to the switching off of the resonance \cite{Kofman:1997yn}. However, the main backreaction for the Higgs field results from the generation of an effective mass due to the self-interaction term $\sim \lambda \langle {\phi}^2 \rangle$, which in fact amplifies the effect since $\lambda<0$ due to curvature induced running if the scale of reheating is large enough, as we explained in the previous section. As an estimate of the significance of the backreaction from the generation of an effective mass from interactions we obtain by using
\ee{\lambda(H) \simeq \lambda_0\, {\rm sign}(\Lambda_I - H)\,,\quad {\rm with} \quad \lambda_0 \approx 0.01\,,
\label{lambda_run}}
with the choice $\mu = H$.
Another backreaction mechanism is given by the potential gravitational significance of the generated particle density. But since the fluctuations scale as $\langle \phi^2\rangle\sim H^2$  so for the Higgs field we have $V(\phi)\sim H^4 \ll H^2M_{\rm pl}^2$, which indicates that gravitational backreaction becomes relevant only for very large variances and hence $\xi$'s.

\begin{figure}
\begin{center}
\includegraphics[width=0.7\textwidth]{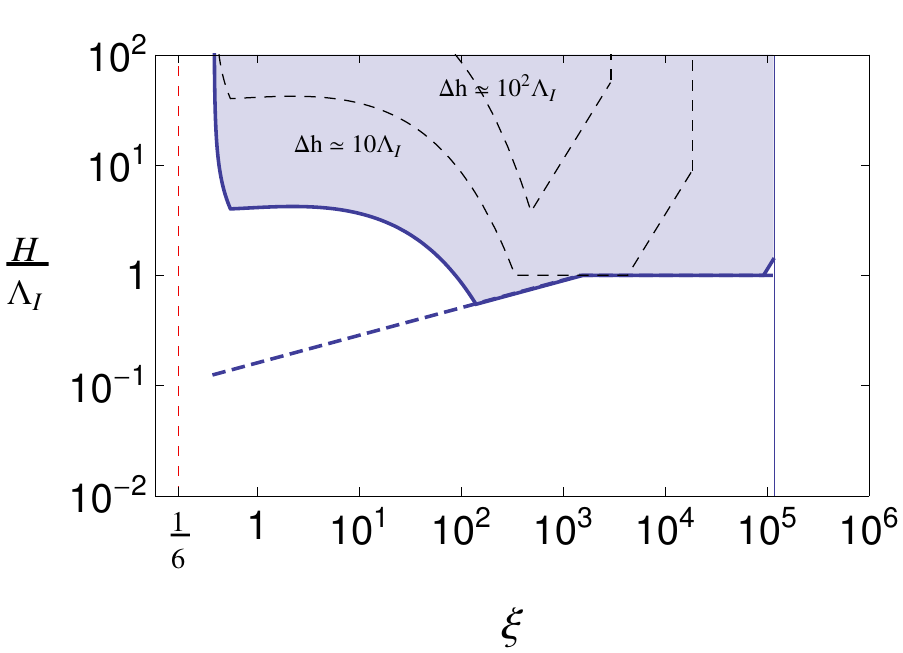}
\caption{\label{fig:regions}The estimated region where there the variance $\sqrt{\langle{\phi}^2\rangle}\equiv \Delta h$ is comparable or larger than  $\Lambda_I$ after the first oscillation for a quadratic inflationary potential. The dashed lines indicate when $\Delta h \gtrsim 10 \Lambda_I$ and $\Delta h \gtrsim 10^2\Lambda_I$ and as the instability scale we have used $\Lambda_I = 10^{-7}M_{\rm pl}$. The region above the dashed line would be largely coloured if particle creation from the subsequent oscillations was taken into account.} 
\end{center}
\end{figure}

In figure \ref{fig:regions} we plot the region where a dangerously large fluctuation is generated for the Higgs field after inflation. We see that for a large scale of inflation/reheating $H\sim \Lambda_I$ even after the first oscillation a vacuum decay can become likely. Here again the crucial ingredient is the curvature induced running of the four-point coupling of the Higgs field, which results in the vanishing of the main backreaction effect, the generation of an effective mass term via the self-interactions. As figure \ref{fig:regions} shows already for $\xi\gtrsim 10$ a vacuum decay becomes likely.

From the analyses of this and the previous section we can draw the important conclusion that a large scale of inflation is compatible with the Standard Model only in a relatively narrow range for the non-minimal coupling
\ee{10^{-2}\lesssim{\xi}\lesssim 10\,,}
which is significantly tighter than the current bound from colliders $|\xi|\lesssim 10^{15}$ \cite{Atkins:2012yn}. From a broader perspective our results indicate that in some instances cosmological consistency can be used as a means for obtaining unprecedented accuracy in bounding parameters of particle physics.
\section{Resonant dark matter generation}
The tachyonic particle production for a non-minimally coupled scalar field as described in the previous section is of course a generic effect not specific to the Standard Model Higgs field. Since after reheating has completed the Universe is dominated by radiation, due to approximate conformal symmetry of the evolution the scalar curvature vanishes, $R=0$. This means that for any scalar field that has no direct coupling to the Standard Model, in the Early Universe there potentially is a mechanism for producing a significant energy-density which quickly after it has been generated becomes decoupled from the evolution of the Universe. Such a behaviour would of course be ideal for a dark matter candidate, as was first discovered in \cite{Markkanen:2015xuw}.

The requirement for a dark matter particle is that it should currently have only weak interactions with the known fields. Another important but less obvious condition is that the dark matter density should consist of only adiabatic perturbations during the formation of the Cosmic Microwave Background. For the dark matter perturbations to be adiabatic, we require that fluctuations in its number density $n_\chi$ exactly match those of the background photons $n_\gamma$
\ea{\delta\bigg(\f{n_\mathbf{\chi}}{n_\mathbf{\gamma}}\bigg)=0\,.\label{dm}}
For a non-minimally coupled scalar field with $\xi \gtrsim 1$ one may easily understand (\ref{dm}) to hold since during inflation the term $\sim \xi R \chi^2$ provides the field a mass and because of this it does not generate a spectrum of long wavelength perturbations unlike the inflaton. During reheating when $R$ oscillates those regions of space that contain a slight over-density of inflatons will also contain a slightly larger $R$ resulting in a greater number of produced dark matter particles. In a way in this mechanism the perturbations in the inflaton field are inherited by the dark matter candidate during reheating. Hence, the dark matter perturbations from tachyonic amplification during reheating are adiabatic, as required.

\begin{figure}
\begin{center}
\vspace{-0.8cm}
\includegraphics[width=0.65\textwidth,trim={3cm 20.55cm  8cm  2cm },clip]{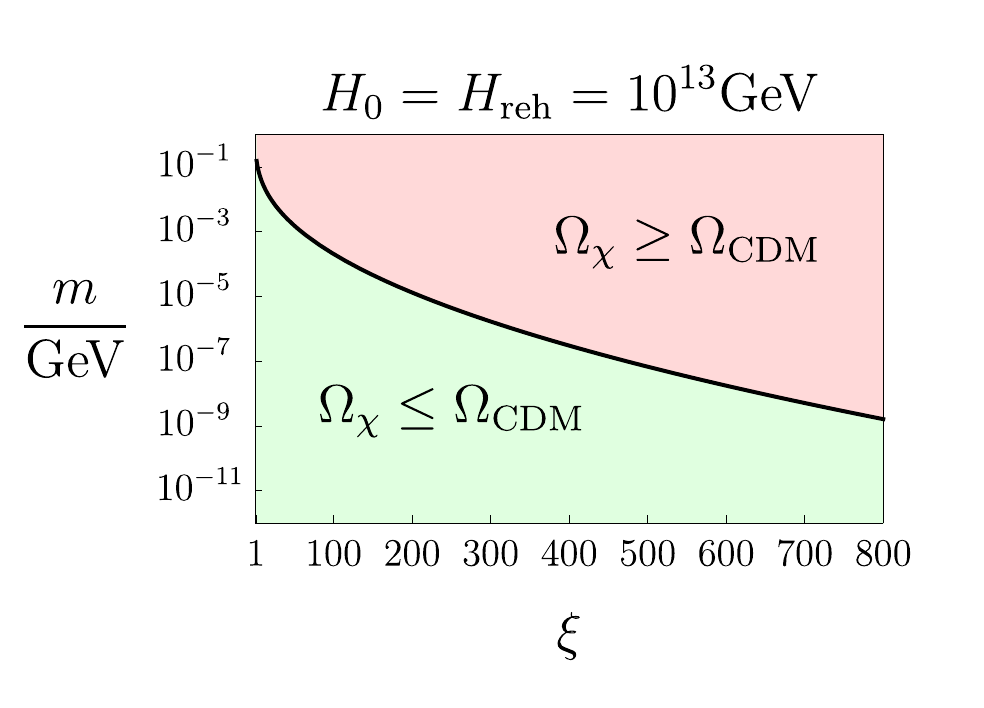}
\end{center}
\vspace{-0.8cm}
\caption{\label{dma}The dark matter abundance resulting from tachyonic amplification due to a non-minimal coupling of a decoupled scalar singlet. We have modelled inflation with $\sim m_\Phi^2\Phi^2$ and assumed that reheating after inflation is instantaneous. $H_{\rm 0}$ and $m$ denote the Hubble rate after the end on inflation and the dark matter mass. Evidently, via this mechanism one may easily generate all of the observed dark matter abundance.} 
\end{figure}

In a simple model where the dark sector contains only the scalar singlet $\chi$ that possesses only a mass term and the non-minimal coupling, an estimate for the produced dark matter abundance can be obtained by duplicating the derivation of the previous section, in particular a lower bound for the number of produced particles can be derived from (\ref{eq:occapp}). 

By using quadratic inflation as a representative model and furthermore assuming reheating to be instantaneous, as shown in \cite{Markkanen:2015xuw} one may write for the dark matter abundance today
\ee{\label{DMabundance}
\frac{\Omega_{\chi}h^2}{0.12} \simeq \xi^{3/8}  \left(\frac{m}{10 \rm GeV}\right) \left(\frac{g_{*,{\rm reh}}}{106.75}\right)^{3/2}\left(\frac{T_{\rm reh}}{10^{15} {\rm GeV}}\right)^3 
\left(\frac{H_{\rm 0}}{H_{\rm reh}}\right)^{3/4} {\rm exp}\left(\frac{2\sqrt{\xi}\Phi_{\rm 0}}{M_{\rm pl}}\right)\,,
}
where $g_{*,{\rm reh}}$ denotes the effective number of relativistic degrees of freedom at the time of reheating, $T_{\rm reh}$ is the temperature of reheating and $H_{\rm 0}$ and $\Phi_{\rm 0}$ denote the Hubble rate and inflaton amplitude after the end on inflation when the coherent oscillations of $\Phi$ start.

In figure \ref{dma} we show the generated dark matter density as a function of the non-minimal coupling and the mass of the scalar singlet. To be conservative we have only included particle creation resulting from the first oscillation. Even with this restriction figure \ref{dma} shows that the tachyonic amplification from the non-minimal term can easily generate all the dark matter abundance as required by experiments.

To conclude we emphasize a further desirable feature of the presented dark matter generation mechanism:  since there is no direct coupling to the inflaton-sector, spoiling the flatness of the inflationary potential is not an issue.
\section{Discussion}
The possibility of a metastable vacuum of the Standard Model can have drastic consequences when the current understanding of the cosmological evolution of the Universe is included in the picture: in the Early Universe where the gravitational dynamics are large a fatal transition to the true vacuum can occur when the scale of inflation is high. The gravitational dynamics during cosmological inflation is not the only dangerous epoch but also during the reheating phase after inflation the instability can materialize. 

It is an observable fact that currently the Higgs field sits firmly in the electroweak vacuum. Since it is unlikely that in the case when the vacuum decay is triggered in the Early Universe the current situation would follow, we can conclude that some stabilizing mechanism, potentially new physics, must exist in order to stabilize the electroweak vacuum if the scale of inflation is high. In this manner the Standard Model vacuum instability in combination with cosmology leads to an indirect probe of beyond the Standard Model physics.

As the works \cite{Herranen:2014cua,Herranen:2015ima} demonstrated, gravity must be incorporated in the quantum dynamics of the Early Universe when calcuating the implications from the vacuum instability. The two effects that are visible only when the curvature of the backround is not neglected are the generation of the non-minimal coupling for the Higgs field and the running of the constants induced by the background curvature. The main conclusion of \cite{Herranen:2014cua,Herranen:2015ima} was that the non-minimal parameter $\xi$ can result in  a stabilizing mechanism during inflation and reheating. Choosing $\xi$ to be non-zero is well-motivated from the field theory point-of-view since $\xi=0$ is not a fixed point of the renormalization group flow. 

During inflation requiring stability gives rise to a lower bound for $\xi$. However interestingly, in the reheating phase having a large $\xi$ results in the formation of a large fluctuation, which indicates that stability during reheating gives rise to an upper bound for $\xi$. When both the inflationary and reheating stability limits are combined one gets that the safe parameter range when stability during inflation and reheating may take place is centered around the conformal point $\xi=1/6$ \cite{Herranen:2014cua,Herranen:2015ima}

\begin{equation}
\mathcal{O}(10^{-2})\lesssim\xi\lesssim\mathcal{O}(10)\,.
\end{equation}

The vacuum instability during reheating is the consequence of a tachyonic amplification due to the presence of the non-minimal coupling to gravity. This effect is generic and occurs for any non-minimally coupled scalar singlet. For a completely decoupled scalar singlet with a non-zero $\xi$-coupling such a tachyonic amplification can act as a viable dark matter generation mechanism \cite{Markkanen:2015xuw}, showing the richness of effects resulting from classical gravity in the Early Universe.
\vskip6pt

\enlargethispage{20pt}




\competing{The author declares that he has no competing interests.}

\funding{The research leading to these results has received funding from the European Research Council under the European Union's Horizon 2020 program (ERC Grant Agreement no. 648680).}

\ack{This article is based on work written in collaboration with Matti Herranen, Sami Nurmi and Arttu Rajantie. Arttu Rajantie and Sami Nurmi also provided helpful insights and suggestions for improving the manuscript.}



\end{document}